# Nonlinear Ionization Dynamics of Hot Dense Plasma Observed in a Laser-Plasma Amplifier

F. Tuitje[1,2], P. Martínez Gil[3], T. Helk[1,2], J. Gautier[4], F. Tissandier[4], J.-P.Goddet[4], A. Guggenmos[5,6], U. Kleineberg[5], S. Sebban[4], E. Oliva[3], C. Spielmann[1,2], and M. Zürch[1,2,7,8]

[1] Institute for Optics and Quantum Electronics, Abbe Center of Photonics, University of Jena, Jena, Germany
[2] Helmholtz Institute Jena, Jena, Germany
[3] Departamento de Ingenı́eria Energética, ETSI Industriales, Universidad Politécnica de Madrid, Instituto de Fusión Nuclear "Guillermo Velarde", Universidad Politecnica de Madrid, Madrid, Spain
[4] Laboratoire d'Optique Appliquée, ENSTA Paris, Ecole Polytechnique, CNRS, Institut Polytechnique de Paris, Palaiseau, France
[5] Department for Physics, Ludwig-Maximilian-University Munich, Garching, Germany
[6] UltraFast Innovations GmbH, Garching, Germany
[7] Fritz Haber Institute of the Max Planck Society, Berlin, Germany
[8] University of California at Berkeley, Department of Chemistry, Berkeley, USA

**From fusion dynamics in stars, to terrestrial lightning events, to new prospects of energy production[1,2] or novel light sources[3–5], hot dense plasmas are of importance for an array of physical phenomena. Due to a plethora of correlations in highly excited matter, direct probing of isolated dynamics remains challenging. Here, the 32.8-nm emission of a high-harmonic seeded laser-plasma amplifier (LPA)[3], using eight-fold ionized Krypton as gain medium, is ptychographically imaged in longitudinal direction in the extreme ultraviolet (XUV). In excellent agreement with *ab initio* spatio-temporal Maxwell-Bloch simulations, an overionization of krypton due to nonlinear laser-plasma interactions is observed. This constitutes the first experimental observation of the laser ion abundance reshaping a laser plasma amplifier. The findings have direct implications for upscaling plasma-based XUV and X-ray sources and allow modeling light-plasma interactions in extreme conditions, similar to those of the early times of the universe, with direct experimental verification.**

Ionized matter constitutes up ~99 % of the observable matter in the universe. Highly ionized plasmas, where the ions are multiply charged and the degree of ionization rises above the sub-percentage regime, are, therefore, an important current research topic[6,7]. Determining important plasma parameters such as electron density and temperature profile in a hot plasma is of widespread importance, for instance in nuclear fusion research, uniform shockwave formation in z-pinch experiments[2] or stable confinement conditions in temperature-based fusion setups[1]. Table-top particle accelerators based on the plasma wake-field effect have the ability to implement high energy research to the lab-scale and pave the way to easy access particle and radiation sources for materials science or medicine where optimized spatio-temporal ionization profiles in the host plasma lead to large accelerating gradients[8]. Furthermore, plasma-based radiation sources provide a wide array of radiation from the visible to the X-ray regime and enable applications from spectroscopy[9] to lithography[10]. The properties of the radiation depend on the generation process of the plasma. Spatially and temporally highly coherent and directed radiation[3–5], as well as diffuse incoherent illumination[11–13] can be achieved. Understanding and classifying the inherent effects taking place in this extreme kind of matter is, therefore, a crucial part of exploring the surrounding nature or enhancing technology. While plasma generation is in most cases straight-forward, analyzing its composition is non-trivial. Due to the plethora of internal processes taking place (excitation, recombination, collision, etc.) and the turbulent nature of the gaseous media, the observation methods have to be highly adapted to the plasma conditions.

Here, the inner dynamics of a laser-pumped plasma channel and the induced spatial ionization structure are resolved with high fidelity. The experiments are performed with a table-top near-infrared (NIR) laser-driven soft-X-ray laser (SXRL) acting as a LPA for a high harmonic generation (HHG) seed pulse to obtain high spatial and temporal coherence and a synchrotron like total flux[3] (see Methods). In this system, eight-fold ionized krypton ions ($Kr^{8+}$), pumped by collisions with free electrons of the plasma, act as amplification medium resembling a nickel-like quasi 3-level laser scheme. This enables a strong amplification in the XUV domain. The amplifying medium, consisting of highly ionized krypton atoms, is generated by an intense laser pulse. Hence, the ionization state is controllable by the laser and gas parameters. The seed, which experiences amplification, is simultaneously acting as a probe of the laser-plasma interaction. Subsequently, ptychographic coherent diffraction imaging is employed to measure the coherent complex-valued emitted radiation field with high spatial resolution. The ptychographic reconstruction of an arbitrary object directly recovers the complex-valued wavefront[14], which allows a backpropagation to the LPA. Four-dimensional Maxwell-Bloch simulations are used to model the amplification of the HHG beam throughout the plasma, i.e. in the forward direction, taking plasma dynamics and inhomogeneities into account. The spatio-temporal simulated LPA outputs are in excellent agreement with the experimental observations. This enables a comprehensive view of the ionization dynamics in the laser-generated plasma and reveals an inner sight of the ionization mechanisms. The results indicate that nonlinear ionization takes place in the amplifier.

**Experimental results**
The LPA gain medium is a plasma of $Kr^{8+}$ ions created via optical field ionization by an intense NIR femtosecond pulse[15]. Amplification of the HHG seed pulse takes place on the $3d^94d$-$3d^94p$ transition of the $Kr^{8+}$ ion at 32.8 nm. Pumping of the population inversion between these two states is ensured by collisions with the hot electrons of the plasma, mainly from the ground state of the $Kr^{8+}$ ions (Fig. 1A). The HHG seed spectrum can be tuned such that one harmonic of the seed pulse resonates with the lasing transition, enabling stimulated emission. During this coherent amplification process, specific characteristics to the laser plasma amplification process depending on the lasing ion and electron density get imprinted on the seed pulse. See Methods for more details.

The seeded LPA is imaged and demagnified onto a sample composed of a regular hole pattern by a system of two spherical mirrors (Fig. 1B, see Methods). The coherently diffracted light is then recorded in transmission geometry. In order to retrieve the complex-valued illumination field, further called *probe*, ptychography is employed[14,16]. In contrast to other wavefront sensing techniques, such as e.g. a Hartmann-Shack sensor, ptychography enables a higher spatial sampling which is important for precise free-space backpropagation. Additionally, using an object with aperture sizes comparable to the focus size, a higher flux on the detector can be achieved reducing the exposure time. Recovering the phase is crucial to enable the backpropagation to the source and, therefore, to analyze the plasma. To achieve a low reconstruction error, the object was scanned on a spiral path with 30 overlapping scanning points (Fig. 1B). For further details see Methods. At each scan position five diffraction patterns were recorded allowing subsequent averaging. The retrieved probe represents the coherent part of the full illumination field (amplitude depicted in Fig. 1C) and shows a diameter of 5.6±0.2 µm at full width at half maximum (FWHM). With an emitting diameter of the plasma channel of 90±10 µm and a demagnification of 10, this leads to 40±10 % of the beam area being spatially coherent, which is substantially higher in comparison to a non-seeded SXRL[4] and similarly to a free-electron Laser[17]. Following the successful retrieval of the complex-valued illumination field in the sample plane, a backpropagation to the exit plane of the plasma channel using the angular spectrum propagation method is performed. See methods and supplementary materials Section S1 for more details.

The experimentally obtained complex-valued exit field of the LPA is depicted in the inset of Figure 2. A dip is observed in the radial intensity profile (Fig. 2, blue solid line). The radial phase profile (Fig. 2, red dashed line) shows a parabolic shape. The increased standard deviation of the phase for larger radii emerges from the near-zero intensity, due to the reconstruction process and the consequently random phases.

**Four-dimensional Maxwell-Bloch Simulations**
In order to understand the complex plasma dynamics that result in the observed LPA output, it is necessary to fully model the propagation of the HHG seed pulse in three spatial dimensions within the LPA. Additionally, due to the ultrafast nature of the seed pulses, the amplification process is non-adiabatic and requires appropriate time-domain modelling.

The amplification in the plasma is modelled with the 3D Maxwell-Bloch code *Dagon*[18]. This code solves the Maxwell wave equation for the electric field in the paraxial form using the slowly varying envelope approximation (SVEA). This equation is enhanced with a constitutive relationship for the polarization and rate equations for the upper and lower level populations of the lasing transitions. These equations are derived from Bloch equations. The temporal evolution of the plasma after pumping is obtained from a collisional-radiative code, *OFIKinRad*[19] and previous particle-in-cell (PIC) modelling[20,21] with the *WAKE-EP*[22] and *Calder-Circ*[23] codes. The plasma waveguide profile was obtained from experimental results[3] and hydrodynamic simulations [21] with the code *ARWEN*[24]. For further information, see Methods.

Figure 3 shows the electron density profile of the plasma waveguide and the $Kr^{8+}$ distribution along the LPA. The NIR and HHG beams propagate from the upper to the lower part of the depicted waveguide. The steep rise on electron density at the bottom part of the figure marks the position of the NIR pump pulse, which creates the lasing ion by optical field ionization of the $Kr^{3+}$ ions composing the waveguide. The lasing ion ($Kr^{8+}$) profile after the NIR pump pulse traverses the amplifier is shown in Fig. 3B. A radial Gaussian profile is assumed for its abundance, with a flat-top region near the central part of the plasma, as given by PIC modelling. Taking the radial profile of the plasma waveguide into account results in a small parabolic structure near the central part of the channel. The central part of the amplifier appears to be overionized, following PIC modelling. Focusing effects increase the intensity of the pump NIR beam in some regions of the amplifier, attaining the threshold to produce higher charged ions. Thus, $Kr^{8+}$ is depleted in these regions and the electron density is further increased. This result fits well to the observed center dip of the experimentally observed exit wave.

The spatio-temporal intensity distribution of the amplified HHG, as given by Maxwell-Bloch modelling, is shown in Fig. 4A. The duration of the pulse is of few hundreds of femtoseconds (< 300 fs), in good agreement with previous experimental and modelling results[3]. The HHG amplified beam presents a rich structure, induced by the amplifier radial profile and its inhomogeneities. Temporal oscillations (Rabi oscillations between $3d^94d$ and $3d^94p$ states) with a period of approximately 80 fs are clearly visible. In addition to this, intensity isocontours have a curved profile, induced by the radial distribution of the plasma waveguide. Instead of a single intensity maximum at the center of the amplifier, two intensity maxima appear at several micrometers away from the central part. The parabolic shape of the plasma channel's electron density along with the overionization in the channel reduce the amplification in the central part resulting in the two-peaked profile and the phase that the experiment revealed in excellent agreement with the simulation, shown in Fig. 4B.

**Summary and Discussion**
In this work, ptychographic imaging successfully employed to retrieve complex-valued illumination functions with high resolution and fidelity, and is applied for the first time to image plasma dynamics in a LPA. While spatial filtering by the amplifier gain is usually expected to clean up the beam profile, here a modulated wavefront is observed, indicating an inhomogeneous distribution of the gain medium. Spatio-temporal calculations unravel the laser-plasma dynamics in the otherwise inaccessible high-density plasma channel. The simulation yields excellent agreement with the experimental observation, validating the parameters and models chosen to reproduce the complex ionization dynamics inside the waveguide. It is found that the observed inhomogeneous amplification is a consequence of the propagation of the required strong optical pump pulse. The accompanying nonlinear ionization results in a heterogeneous electron density, which directly correlates to the ionization degree of krypton. Overionization in the center of the channel causes the $Kr^{8+}$ abundance to decreases locally to 20 % of neutral density and results in an inhomogeneous amplification, which is imprinted on the output wave of the LPA. More generally, the results indicate the limitations of upscaling LPA and related SXRL technologies based on optical field ionization and state boundaries for laser-based generation of hot dense plasmas with certain ion compositions. Furthermore, the observations reported here show the importance of four-dimensional modeling of the laser-plasma interaction especially in highly ionized plasmas that can lead to substantial reshaping of all involved pulses. To our knowledge, this is the first observation where *ab-initio* modeling predicts correctly the amplification in a LPA. The experimental validation of the used models holds great promise to employ these numerical methods to predict future LPA and SXRL schemes. Finally, the observed Rabi oscillations in the LPA lead to strong modification of the temporal structure of seed pulses. This approach of modulating XUV pulses could open possibilities for e.g. qubit state control[25] or tracking of electronic[26] or molecular[27] dynamics and is highly stable due to the fixed transition dipole moment.

**Methods**
*Experimental setup*
The experiment was performed at Laboratoire d'Optique Appliquée (LOA) using the *Salle jaune* Ti:Sapphire laser system[3] able to deliver three independently-compressed multiterawatt femtosecond pulses. The target is a high-density krypton gas jet equipped with a 5 mm–long rectangular nozzle. It is pumped by a 1.4 J, 30 fs pulse focused by a 0.8 m focal length spherical mirror. Since the plasma electron density typically ranges from a few $10^{19}$ to $10^{20}$ cm$^{-3}$, the pump pulse cannot propagate over more than a few hundreds of microns, so an optically-preformed plasma waveguide was implemented. A 100 mJ, 30 fs "ignitor" pulse followed by a 700 mJ, 600ps "heater" pulse are focused into the krypton jet using an axicon lens. After hydrodynamic expansion of the plasma, a 5 mm long channel suitable to guide the main pump pulse is achieved. After passing of the pump pulse, the plasma contains a high fraction of $Kr^{8+}$ ions as well as the hot electrons needed to pump the population inversion. The amplifier is seeded by a HHG pulse created by focusing a 15 mJ pulse into an argon gas cell. The HHG source is imaged onto the amplifier using a grazing-incidence toroidal mirror and the 25$^{th}$ harmonic can be tuned to the lasing transition by chirping the driver beam. Due to collisional ionization of the lasing ions, the gain has a short lifetime and is strongly peaked[28]. The time delay between the creation of the amplifier and seeding is set at the experimentally obtained value of 1.2 ps to match that peak for strongest amplification. After exiting the LPA and passing two aluminum filters blocking residual NIR light, the XUV pulses are focused onto the sample by two spherical multilayer mirrors with 5 m (mirror 2) and 0.5 m (mirror 1) focal length, respectively. This telescope results in a demagnification of 10. In order to reduce aberration and matching the multilayer conditions, the angle of incidence on the curved mirrors was minimized to 4°. To implement ptychographic scanning, the sample is mounted on a 3D positioner with a repeatability of 50 nm over the complete travel range of 30 mm. The intensity distribution of diffraction patterns was recorded with a 2048×2048 pixel CCD detector with 16 bit dynamic range and 27.8 mm diagonal size (Andor iKon-L SO). The camera was cooled down to -50°C to reduce electronic noise. Since the sample is 55 mm away from the camera, the numerical aperture of our system (0.25) sets the resolution limit to ~65 nm.

*Sample*
The sample, mounted in the focal plane of mirror 1, consists of a gold coated carbon grid with an overall thickness of 50 nm with a regular hole pattern with 1 µm hole diameter and 2 µm pitch, see Fig. 1B.

*Ptychographic reconstruction*
For retrieving the complex-valued wavefront at the plane of the sample, a ptychographic approach was employed. There, the sample was scanned with respect to the XUV beam in a spiral pattern, to avoid so-called grid-pathology artifacts[14], with 30 scan points in total. For a successful reconstruction, the overlap between neighboring scan points has to be fairly high. In consideration of the total amount of exposure, an overlap of 90% was chosen. To derivate a sufficient scan map, a rough estimate of the coherent focal spot size is necessary. A reconstruction of a single diffraction pattern of the sample using the Hybrid-Input-Output algorithm[29] with a feedback parameter β = 0.9 was used to evaluate the rough size of the spot by counting the reconstructed holes resulting in 5.6±0.2 µm FWHM (see Supplementary S3). Using the ePIE algorithm[16] with 3000 iterations and a 6-µm diameter flat-top initial probe, the coherent complex-valued object and illumination functions were retrieved with a resolution of roughly 200 nm. The feedback parameters were chosen to α = β = 0.9. A delay between object and probe reconstruction of 10 iterations was introduced to avoid artifacts from oscillations between object and probe field. To compensate beam drifts and variations, the measurement and reconstruction was repeated five times and averaged.

**Backpropagation**
The used optics and beam path length allows a backpropagation to the exit wave of the plasma channel using the angular spectrum propagation method (see supplementary materials section S1). Here, the mirrors were applied as phase shifts and the incident angles were taken into account to compensate possible coma aberration. Due to the difference between the field-of-views (FOV) of probe field (220 µm) and exit wave (2200 µm), the FOV was adapted by zero padding and oversampling during the propagation to ensure a sufficient bandwidth. To allocate the exit wave in the last part of propagation from mirror 2 to source, the beam was propagated in slices of 10 mm to check the position of the focal spot. Assuming the exit waves location matching the focal spot of mirror 2, the backpropagation ends if the exit wave reaches its most compact form.

**Simulations**
The creation of the plasma channel was modelled with the 2D radiative hydrodynamics code *ARWEN*[29]. Particle-In-Cell codes (*WAKE-EP*[30] and *CalderCirc*[31]) were used to model the propagation of the NIR pulse throughout the plasma channel[29]. The resulting 3D electron density and $Kr^{8+}$ abundance was fed to our 3D Maxwell-Bloch code *Dagon*[27]. With this code we modelled the amplification of a high order harmonic throughout a 2 mm inhomogeneous plasma amplifier. This is the effective amplification length as given by our PIC simulations. Since our Maxwell-Bloch model gives the complex-valued electric field of the amplified pulse, its intensity and phase profiles can be compared directly to propagated ptychographic measurements.

**Acknowledgements**: The research leading to these results has received funding from the European Community's Horizon 2020 research and innovation program under grant agreement n° 654148 (LASERLAB EUROPE). M.Z. acknowledges support by the Max Planck Society (Max Planck Research Group) and the Federal Ministry of Education and Research (BMBF) under "Make our Planet Great Again – German Research Initiative" (Grant No. 57427209 "QUESTforENERGY") implemented by DAAD. F.T. acknowledges support by the Federal State of Thuringia and the European Social Fund (ESF) Project 2018 FGR 0080. P. M. G. and E. O. acknowledge support from the European Community's Horizon 2020 research and innovation program under grant agreement 665207, project VOXEL and the Universidad Politécnica de Madrid, project DERKETA. This work is supported by "Investissements d'Avenir" Labex PALM (ANR-10-LABX-0039-PALM).

**Competing Interests:** The authors declare that they have no competing financial interests.

**Correspondence:** Correspondence and requests for materials should be addressed to F. Tuitje (frederik.tuitje@uni-jena.de), E. Oliva (eduardo.oliva@upm.es), and M. Zuerch (mwz@berkeley.edu).

**Author contributions**: F.T., T.H., J.G., F.T., J.P.G., S.S. and M.Z. performed the experiments. F. T., T.H. and M. Z. analysed the data. A.G. and U.K. designed and built the XUV multilayer optics. P. M. G and E. O. performed the simulations. M. Z. conceived the experiment. C. S. and M. Z. supervised the project. All authors contributed to the manuscript.

**References**
1. Helander, P. *et al.* Stellarator and tokamak plasmas: a comparison. *Plasma Phys. Control. Fusion* **54**, 124009 (2012).
2. Li, Z. *et al.* Experimental investigation of Z-pinch radiation source for indirect drive inertial confinement fusion. *Matter Radiat. Extrem.* **4**, 046201 (2019).
3. Depresseux, A. *et al.* Table-top femtosecond soft X-ray laser by collisional ionization gating. *Nat. Photonics* **9**, 817–821 (2015).

4. Zürch, M. *et al.* Transverse Coherence Limited Coherent Diffraction Imaging using a Molybdenum Soft X-ray Laser Pumped at Moderate Pump Energies. *Sci. Rep.* **7**, 5314 (2017).
5. Spielmann, Ch. *et al.* Generation of Coherent X-rays in the Water Window Using 5-Femtosecond Laser Pulses. *Science* **278**, 661 (1997).
6. Cayzac, W. *et al.* Experimental discrimination of ion stopping models near the Bragg peak in highly ionized matter. *Nat. Commun.* **8**, 1–7 (2017).
7. Vinko, S. M. *et al.* Investigation of femtosecond collisional ionization rates in a solid-density aluminium plasma. *Nat. Commun.* **6**, 1–7 (2015).
8. Joshi, C. *et al.* Ultrahigh gradient particle acceleration by intense laser-driven plasma density waves. *Nature* **311**, 525–529 (1984).
9. Sullivan, J. V. & Walsh, A. High intensity hollow-cathode lamps. *Spectrochim. Acta* **21**, 721–726 (1965).
10. Wagner, C. & Harned, N. Lithography gets extreme. *Nat. Photonics* **4**, 24–26 (2010).
11. Legall, H. *et al.* Compact x-ray microscope for the water window based on a high brightness laser plasma source. *Opt. Express* **20**, 18362–18369 (2012).
12. Silfvast, W. T. Intense EUV incoherent plasma sources for EUV lithography and other applications. *IEEE J. Quantum Electron.* **35**, 700–708 (1999).
13. Schmitz, C. *et al.* Compact extreme ultraviolet source for laboratory-based photoemission spectromicroscopy. *Appl. Phys. Lett.* **108**, 234101 (2016).
14. Thibault, P., Dierolf, M., Bunk, O., Menzel, A. & Pfeiffer, F. Probe retrieval in ptychographic coherent diffractive imaging. *Ultramicroscopy* **109**, 338–343 (2009).
15. Sebban, S. *et al.* Demonstration of a Ni-Like Kr Optical-Field-Ionization Collisional Soft X-Ray Laser at 32.8 nm. *Phys Rev Lett* **89**, 253901 (2002).
16. Maiden, A. M. & Rodenburg, J. M. An improved ptychographical phase retrieval algorithm for diffractive imaging. *Ultramicroscopy* **109**, 1256–1262 (2009).
17. Vartanyants, I. A. *et al.* Coherence Properties of Individual Femtosecond Pulses of an X-Ray Free-Electron Laser. *Phys. Rev. Lett.* **107**, 144801 (2011).
18. Oliva, E. *et al.* DAGON: a 3D Maxwell-Bloch code. in (eds. Klisnick, A. & Menoni, C. S.) 1024303 (2017). doi:10.1117/12.2265044.
19. Cros, B. *et al.* Characterization of the collisionally pumped optical-field-ionized soft-x-ray laser at 41.8 nm driven in capillary tubes. *Phys. Rev. A* **73**, 033801 (2006).
20. Oliva, E. *et al.* Self-regulated propagation of intense infrared pulses in elongated soft-x-ray plasma amplifiers. *Phys. Rev. A* **92**, 023848 (2015).
21. Oliva, E. *et al.* Hydrodynamic evolution of plasma waveguides for soft-x-ray amplifiers. *Phys. Rev. E* **97**, 023203 (2018).
22. Paradkar, B. S., Cros, B., Mora, P. & Maynard, G. Numerical modeling of multi-GeV laser wakefield electron acceleration inside a dielectric capillary tube. *Phys. Plasmas* **20**, 083120 (2013).
23. Lifschitz, A. F. *et al.* Particle-in-Cell modelling of laser–plasma interaction using Fourier decomposition. *J. Comput. Phys.* **228**, 1803–1814 (2009).
24. Ogando, F. & Velarde, P. Development of a radiation transport fluid dynamic code under AMR scheme. *J. Quant. Spectrosc. Radiat. Transf.* **71**, 541–550 (2001).
25. Carlström, S., Mauritsson, J., Schafer, K. J., L'Huillier, A. & Gisselbrecht, M. Quantum coherence in photo-ionisation with tailored XUV pulses. *J. Phys. B At. Mol. Opt. Phys.* **51**, 015201 (2017).
26. Wituschek, A. *et al.* Tracking attosecond electronic coherences using phase-manipulated extreme ultraviolet pulses. *Nat. Commun.* **11**, 1–7 (2020).
27. Loriot, V. *et al.* Resolving XUV induced femtosecond and attosecond dynamics in polyatomic molecules with a compact attosecond beamline. *J. Phys. Conf. Ser.* **635**, 012006 (2015).
28. Zeitoun, P. *et al.* A high-intensity highly coherent soft X-ray femtosecond laser seeded by a high harmonic beam. *Nature* **431**, 426–429 (2004).
29. Fienup, J. Reconstruction of a complex-valued object from the modulus of its Fourier transform using a support contraint. *Opt. Soc. Am.* **4**, 521–553 (1987).
30. Jaeglé, P. *Coherent Sources of XUV Radiation*. (Springer International Publishing, 2006).

**Figures**

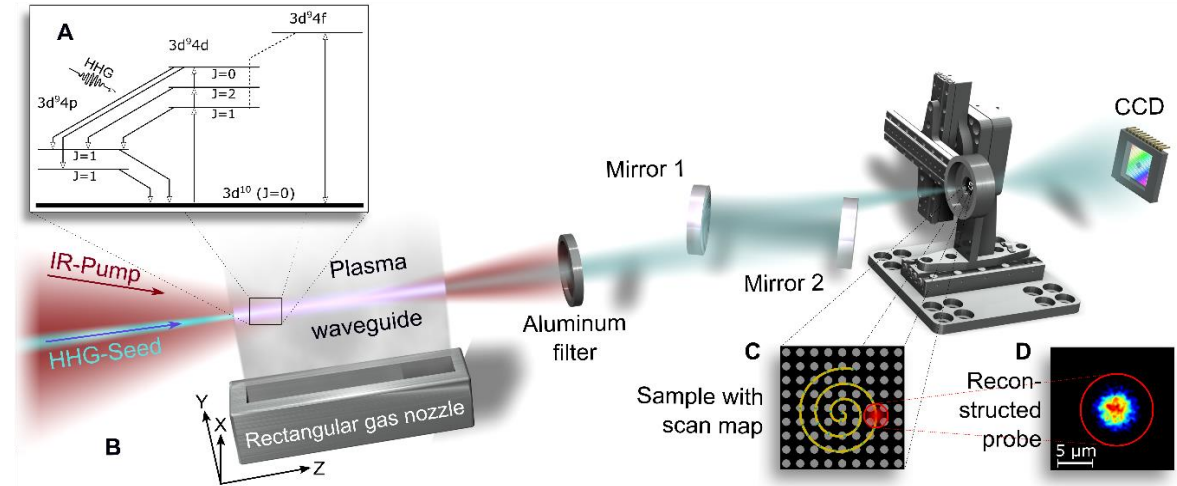

**Fig. 1: Experimental setup and operation scheme of laser-plasma amplifier with diagnostics. A:** A series of infrared pump beams (see Methods) create a plasma waveguide, excite nickel-like $Kr^{8+}$ and create population inversion of the $3d^94d$ state, forming a LPA. Here, only dipole-allowed transitions involved in the amplification process are shown (inset level scheme[30]). The HHG seed at 32.8 nm wavelength is coupled into the plasma channel and is amplified by stimulated emission (4d - 4p transition in $Kr^{8+}$). **B**: Schematic setup of the experiment. **C**: The emitted radiation is refocused using multilayer mirrors onto the sample consisting of a regular hole pattern (**C** depicts SEM image). Ptychography is performed using a spiral scan pattern (indicated in yellow) with a CCD recording a coherent diffraction pattern at each scan point. The relation between scan map and probe size is marked with red circles **D:** The recorded diffraction patterns using ptychography can be reconstructed to retrieve the coherent complex-valued illumination function (probe) (**D** depicts the amplitude of the reconstructed illumination function).

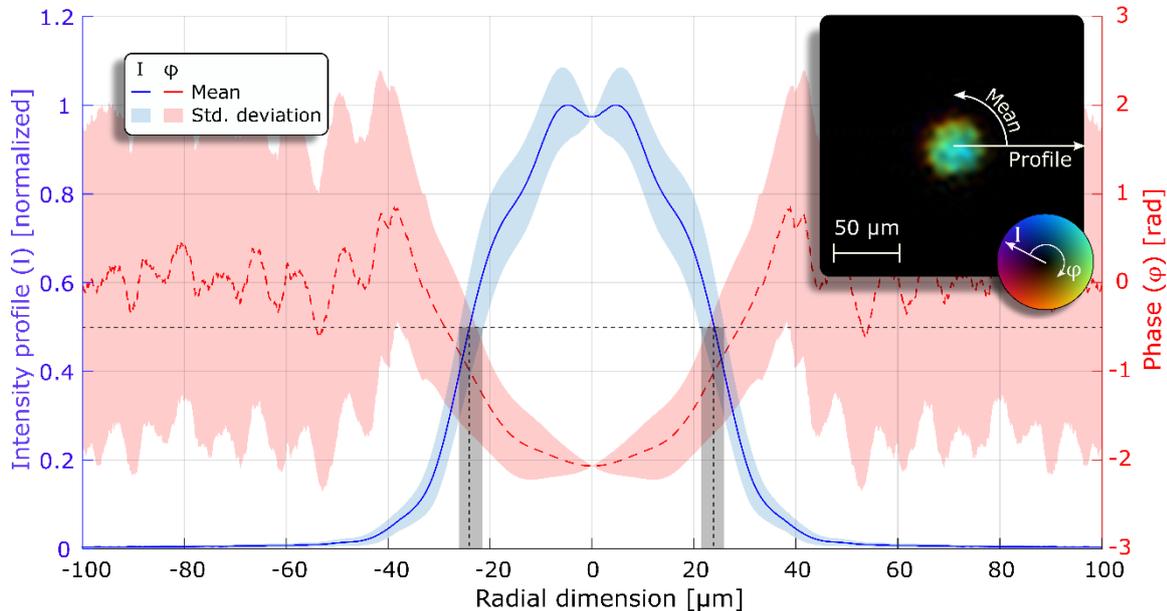

**Fig. 2: Reconstructed exit field of the laser-plasma amplifier.** The complex-valued retrieved exit field of the LPA is pictured in the inset. Here, hue represents phase and brightness represents intensity. The radial profile of the intensity shows a Gaussian-like profile with a dip in the center. Simulations indicate that an overionized zone in the LPA leads to decreased amplification in the center of the channel. The phase profile shows a parabolic shape caused by the radially decreasing refraction index. Note: the high standard deviation of the phase above 40 µm radius arises from the low intensity and the corresponding random phases during the reconstruction. The diameter of 52±5 µm (FWHM) of the exit field is marked with the black dashed vertical lines, where the gray bar represents the error.

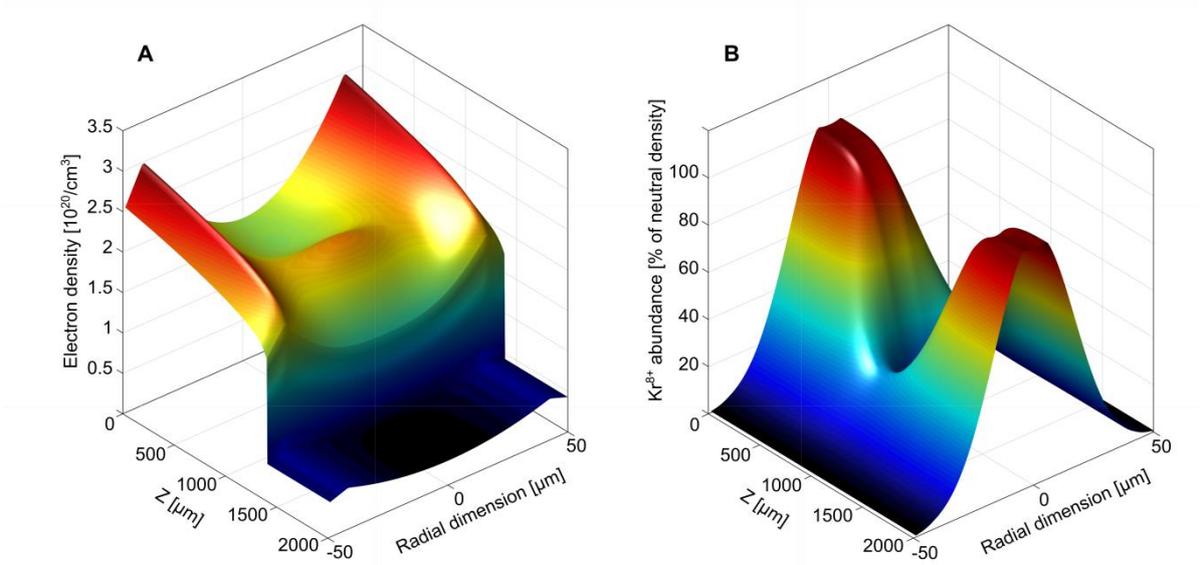

**Fig. 3: Spatial distribution of electrons and lasing ions in the LPA following the NIR pump pulse.**
**A:** Electron density profile of the plasma waveguide after propagation of the pump pulse to z=1200 µm.
**B:** Lasing ion ($Kr^{8+}$) abundance in the LPA in percent of neutral density after complete propagation through the channel. The lasing ion is depleted at Z=1000 µm at the radial center due to overionization. Thus, the electron density profile shows a corresponding peak in this region. Furthermore, **B** point out a groove of decreased ion abundance for r=0 µm, resulting in an attenuated amplification, explaining the dip in intensity observed in the experiment (Fig. 2).

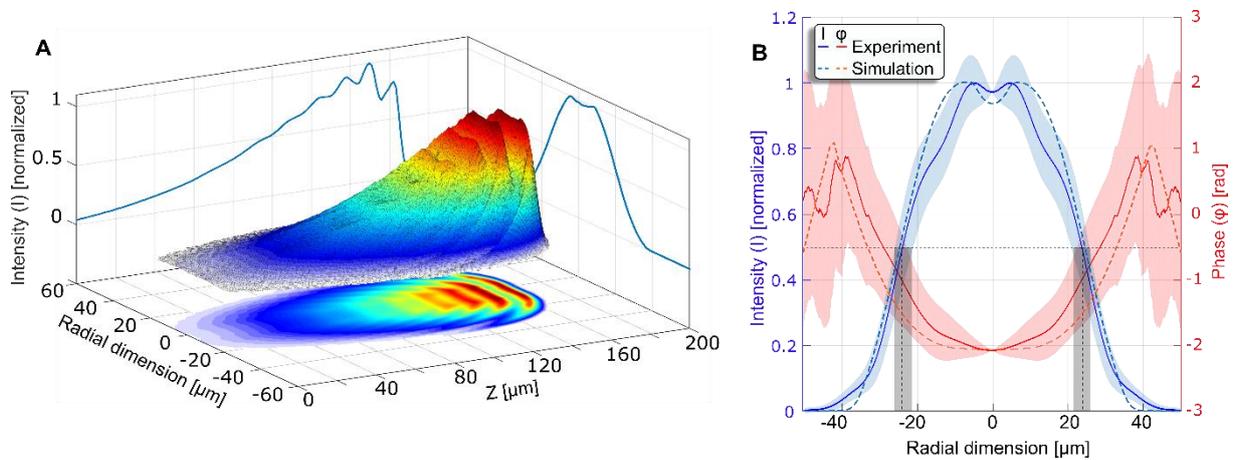

**Fig. 4: Spatio-temporal intensity profile of the amplified HHG pulse and comparison to experiment. A** The beam shows a rich structure with temporal (Rabi) oscillations. The curved iso-intensity contours reveal two intensity maxima that are not located at the central part of the amplifier. All these structures are induced by the plasma waveguide inhomogeneous profile and the lasing ion abundance, through its radial profile and the depletion of lasing ions in the central part of the amplifier. **B:** the numerically accumulated intensity and phase shows excellent agreement with the experiment. Errors of experimental data are shown in pale colors and the black bars indicate the diameter.

# Supplementary Information

# Nonlinear Ionization Dynamics of Hot Dense Matter Observed in a Seeded Soft X-Ray Laser

F. Tuitje[1,2], P. Martínez Gil[3], T. Helk[1,2], J. Gautier[4], F. Tissandier[4], J.-P.Goddet[4],
A. Guggenmos[5,6], U. Kleineberg[5], S. Sebban[4], E. Oliva[3], C. Spielmann[1,2], and M. Zürch[1,2,7,8]

[1] Institute for Optics and Quantum Electronics, Abbe Center of Photonics, University of Jena, Jena, Germany
[2] Helmholtz Institute Jena, Jena, Germany
[3] Departamento de Ingenéria Energética, ETSI Industriales, Universidad Politécnica de Madrid, Instituto de Fusión Nuclear "Guillermo Velarde", Universidad Politecnica de Madrid, Madrid, Spain
[4] Laboratoire d'optique appliquée, ENSTA-ParisTech, Palaiseau, France
[5] Department for Physics, Ludwig-Maximilian-University Munich, Garching, Germany
[6] UltraFast Innovations GmbH, Garching, Germany
[7] Fritz Haber Institute of the Max Planck Society, Berlin, Germany
[8] University of California at Berkeley, Department of Chemistry, Berkeley, USA

## S1 Angular spectrum propagation of the reconstructed probe

To propagate the reconstructed complex-valued illumination function back to its source, a bandwidth limited angular spectrum propagator was used[1]. The mirrors were applied as a curved phase shift overlaid with a tilted plane under a reflection angle $\alpha$

$$\Phi_M = \exp\left(i\frac{2\pi}{\lambda}\sqrt{r^2 - u^2 - v^2}\right) \exp\left(i\frac{2\pi}{\lambda}\tan(\alpha)u\right) \quad (1)$$

onto the propagated field $\Phi_{in}$ via an elementwise complex Hadamard product

$$\Phi_{out} = \mathfrak{P}(\Phi_{in}, d) \odot \Phi_M \quad (2)$$

$\mathfrak{P}$ assign the angular spectrum propagator to

$$\mathfrak{P}(\Phi_{out}, d) = \mathfrak{F}^{-1}\left(\mathfrak{F}(\Phi_{in}) \odot \exp\left(i\frac{2\pi}{\lambda}d + i\pi\lambda d(u^2 + v^2)\right)\right) \quad (3)$$

with $\mathfrak{F}$ and $\mathfrak{F}^{-1}$ as the Fourier resp. inverse Fourier transform, $\lambda$ as the wavelength, $d$ as the propagation distance and $u, v$ as the pixel coordinates of the fields with a centered origin. Further propagation via mirror 2 to the position of the exit wave completes the path of propagation to:

$$\Phi_{exit} = \mathfrak{P}(\mathfrak{P}(\mathfrak{P}(\Phi_{probe}, d_1) \odot \Phi_{M1}, d_2) \odot \Phi_{M2}, d_3) \quad (4)$$

Due to the increasing beam waist with propagation to the first mirror, the probe field $\Phi_{in}$ was zero-padded to increase the field-of-view to 1 cm to avoid information loss and aliasing.

## S2 Evaluation of the exit wave's position

The actual position of the exit wave may not exactly agree with the focal distance $d_3$ of mirror 2 due to experimental variances. Hence, the numerical distance were slightly varied to find the focus within an offset distance $\delta = 1m$, varying from $d_3 - \frac{\delta}{2}$ to $d_3 + \frac{\delta}{2}$. By extracting equidistant slices of $\Phi_{exit}$ in $d_3 \pm \frac{\delta}{2}$ for every 10 mm, a focal cross section was determined.

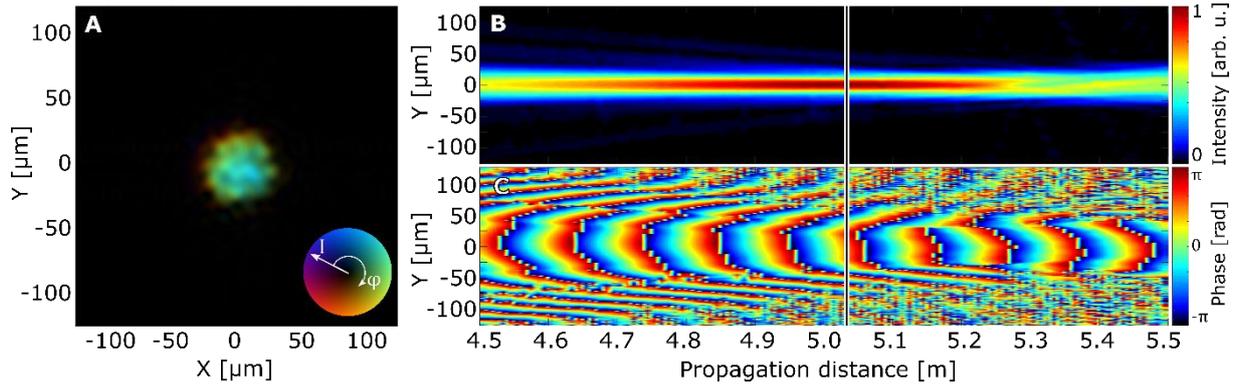

**Fig. 1: Focal cross section around $d_3$.** The last propagation $d_3$ to the exit wave $\Phi_{exit}$ (**A**) was varied with $\delta = 1\,m$ to find the physical focus via an intensity (**B**) and phase (**C**) cross section. Here the flip of sign of phase curvature (white/black line in **B**,**C**) was used as a criterion together with the position of highest intensity in (**B**). In (**C**), the visible period of ~10 cm represents not the wavelength, rather it is formed by aliasing due to the slice-to-slice distance.

Figure 1**B**,**C** show the focal point in phase and intensity of the numerical propagation after mirror 2, which is ~35 mm shifted along the beam direction with respect to the mirror defined focal length of 5 m. The final exit wave is shown in Figure 1**A** and is used for the comparison to the simulation.

## S3 Focus size and beam stability

Creating the ptychographic scan map requires a rough estimate of the coherent spot size in the sample plane[2,3]. We used a CDI reconstruction of a single-shot diffraction pattern of the sample (Quantifoil Micro Tools GmbH, Quantifoil® R1/2) to investigate the spot size as the reconstruction can be considered as the product between the illuminating field and the sample itself. After the reconstruction, a 2D Gaussian function can be fitted to the reconstructed structure, as seen in Fig. 2. In this way, the coherent part of the total illumination in the sample plane can be read out to 5.6 ± 0.2 µm FWHM.

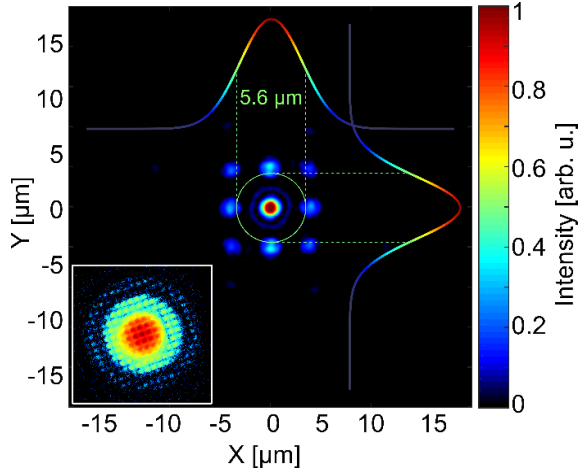

**Fig. 2: Reconstructed periodic hole sample with corresponding diffraction pattern (inset).** By fitting a 2D gaussian function to the reconstruction, the coherent width (FWHM) of the illumination function is estimated to be ~5.6 µm. The aspect ratio is ~1.

Ptychography without sophisticated techniques as e.g. mixed state reconstruction[4] or map refinement[5], requires spatially stable beams and low shot-to-shot variations of the wave fronts. We analyzed variations between consecutive shots by recording 100 single-shot diffraction patterns of an isolated object with an overall size of ~5 µm. Correlating the shots to a far-field simulation of the object revealing possible beam jitter. Here, a limit of correlation of 0.75 was set to separate stable diffraction patterns from unstable ones. ~40 % of the recorded pattern can be considered as stable and, therefore, ~40% of shots can be considered as jitter free. For this reason, the measurement of the main dataset was repeated several times to ensure stable results for every scanning position.

## S4 Discussion of other plasma detection methods

The commonly used plasma interferometry technique can assess the electron density of a medium by measuring the phase shift undergone by a well-timed coherent probe beam propagating in the plasma. After performing an Abel inversion of the interfered probe beam with an undisturbed reference beam, the refraction index and electron density can be reconstructed[6–8]. The smallest resolvable features are limited by the size and contrast of interference fringes. Shadowgraphy, where light traverses the plasma and is recorded afterwards, requires a less sophisticated setup but is more limited regarding its quantitative output. Out of the resulting contrast, the refraction index and absorption can be inferred[9]. This method also requires an Abel inversion and is, therefore, limited to radial-symmetric homogenous plasmas. A full characterization of the ionization states in their spatial distribution is hardly realizable[10].

## References

1. Delen, N. & Hooker, B. Free-space beam propagation between arbitrarily oriented planes based on full diffraction theory: a fast Fourier transform approach. *JOSA A* **15**, 857–867 (1998).
2. Maiden, A. M. & Rodenburg, J. M. An improved ptychographical phase retrieval algorithm for diffractive imaging. *Ultramicroscopy* **109**, 1256–1262 (2009).
3. Bunk, O.; Dierolf, M.; Kynde, S.; Johnson, L.; Marti, O.; Pfeiffer, F. Influence of the overlap parameter on the convergence of the ptychographical iterative engine. *Ultramicroscopy* **108**, 481–487 (2007).
4. Thibault, P. & Menzel, A. Reconstructing state mixtures from diffraction measurements. *Nature* **494**, 68–71 (2013).
5. Dwivedi, P., Konijnenberg, A. P., Pereira, S. F. & Urbach, H. P. Lateral position correction in ptychography using the gradient of intensity patterns. *Ultramicroscopy* **192**, 29–36 (2018).
6. Bukin, V. V., Garnov, S. V., Malyutin, A. A. & Strelkov, V. V. Interferometric diagnostics of femtosecond laser microplasma in gases. *Phys. Wave Phenom.* **20**, 91–106 (2012).
7. Cao, L. *et al.* Space-time characterization of laser plasma interactions in the warm dense matter regime. *Laser Part Beams 2520072239-244* **25**, (2008).
8. Sävert, A. *et al.* Direct Observation of the Injection Dynamics of a Laser Wakefield Accelerator Using Few-Femtosecond Shadowgraphy. *Phys. Rev. Lett.* **115**, 055002 (2015).
9. Gopal, A., Minardi, S. & Tatarakis, M. Quantitative two-dimensional shadowgraphic method for high-sensitivity density measurement of under-critical laser plasmas. *Opt. Lett.* **32**, 1238–1240 (2007).
10. Schnell, M. *et al.* Deducing the Electron-Beam Diameter in a Laser-Plasma Accelerator Using X-Ray Betatron Radiation. *Phys. Rev. Lett.* **108**, 075001 (2012).